\documentclass[prd,reprint,showpacs,showkeys]{revtex4}
\usepackage{graphics}
\usepackage{eurosym}
\usepackage{amsfonts}
\usepackage{amssymb}
\usepackage{amsmath}
\usepackage{graphicx}
\usepackage[font={footnotesize,it}]{caption}
\usepackage[colorlinks=false,
            linkcolor=red,
            urlcolor=red,
            citecolor=blue]{hyperref}

\usepackage{color}

\begin{document}

\title{Exact solution of anisotropic compact stars via. mass function}

\author{S. K. Maurya}
\email{sunil@unizwa.edu.om}
\affiliation{Department of Mathematical and Physical Sciences,
College of Arts and Science, University of Nizwa, Nizwa, Sultanate
of Oman}

\author{ Ayan Banerjee}
\email{ayan_7575@yahoo.co.in}
\affiliation{Department of Mathematics, Jadavpur University, Kolkata 700032, West Bengal, India}

\author{ Y. K. Gupta}
\email{kumar001947@gmail.com}
\affiliation{Department of Mathematics, Raj Kumar Goel Institute of Technology, Ghaziabad, 201003, U.P., India}

\date{\today }

\begin{abstract}
The interest in studying relativistic compact objects play an important
role in modern astrophysics with an aim to understand several astrophysical
issues. It is therefore natural to ask for internal structure and physical
properties of specific classes of compact star for astrophysical observable,
and we obtain a class of new relativistic solutions with anisotropic distribution
of matter for compact stars. More specifically, stellar models, described by the
anisotropic fluid, establish a relation between metric potentials and generating
a specific form of mass function are explicitly constructed within the framework
of General Relativity. New solutions that can be used to model compact objects which adequately describe
compact strange star candidates like SMC X-1, Her X-1 and 4U 1538-52, with observational
data taken from Gangopadhyay \emph{et al.} \cite{Gangopadhyay}. As a possible astrophysical
application the obtained solution could explain the physics of  self-gravitating objects,
might be useful for strong-field regimes where data are currently inadequate.
\end{abstract}


\maketitle

\section*{I.~~~Introduction}
Over the past few years, there has been growing interest in static spherically symmetric compact object consistent with general relativity which are most often observed as pulsars, spinning stars with strong magnetic fields. However, in spite of the fact, it is important to know the exact composition and nature of particle interactions, which allow us to completely describe them in terms of their mass, spin angular
momentum and charge. In particular, observations of compact stars are considered primary targets of the forthcoming field of gravitational wave  astronomy. In the recent times we have experimental evidence that such objects do exist from observations with very high densities \cite{Lattimer}, and some of the compact objects like X-ray pulsar Her X-1, X-ray burster 4U 1820-30, X-ray sources 4U 1728-34, PSR 0943+10, millisecond pulsar SAX J 1808.4-3658, and RX J185635-3754 strongly favour the possibility that they could actually be strange stars. In other words, there is no any strong evidence to make conclusion/ understood the mechanism about compact objects.  In spite of this drawback, the internal composition and consequent geometry of such objects are still considered an open question in the scientific community. From a theoretical point of view, such compact objects are composed of a perfect fluid \cite{Ivanov}.  Generally, polytropic equation of state (EOS) in the form P = k $\rho^{\gamma}$ and bag model
\cite{Witten,Glendenning} have widely used to describe a white dwarf or a less compact star \cite{Shapiro}.
An example, Herrera and Barreto \cite{Barreto} carry out a general study on polytropic general relativistic stars with anisotropic pressure whereas Lai and Xu \cite{Lai} have studied polytropic quark star model.

  Though, interior of a star is an important astrophysical question and hence it is pertinent
to construct relativistic models by assuming anisotropy fluid distribution.
The theoretical study of the influence of anisotropic compact objects was first initiated by Bowers and Liang
\cite{Bowers} and  another investigation led be Ruderman \cite{Ruderman} showed that nuclear matter may have anisotropic features at least in certain very high density ranges $10^{15}$ g/$cm^3$, where the nuclear interaction
must be treated relativistically. Later on, several  models (see
for example, \cite{Gleiser,Dev2003,Herrera2002}) have been proposed in this direction.
The procedure has been found to be interesting and useful on the physical grounds that
anisotropy affects the critical mass, mass-radius relation and stability of highly compact relativistic bodies.
It is also well-established fact that a magnetic field acting on a Fermi gas produces pressure
anisotropy was discussed recently in \cite{Chaichian,Martinez,Ferrer}. Recently, Maurya and Gupta in \cite{Maurya2015a,Maurya2016a} have studied charged anisotropic stars whereas without charged solution has been analysed in \cite{Hossein,Mehedi2012}.
The role of anisotropy, with the linear equation of state  was pursed by Mak and Harko \cite{Mak(2002)}, and
extending the work by Sharma and Maharaj \cite{Sharma} obtained an exact analytical solutions assuming a particular form of mass function. In yet another paper, Victor Varela et al \cite{Varela}
studied Charged anisotropic matter with linear or nonlinear equation of state .

However, it has recently been proposed that an arbitrary d  dimensional
(pseudo-) Riemannian space can always be locally embedded  into any Riemann
space of dimension N $\geq$ d(d + 1)/2. Riemann's seminal work in 1868 \cite{Riemann} had inspired
Schl$\ddot{a}$fli that how one can locally embed such manifolds in Euclidean space \cite{Schlafli}.
He discussed the local form of the embedding problem and conjectured that
maximum number of extra dimensions that can embedded as a submanifold of a Euclidean space $E_N$ with N= $\frac{1}{2}n(n+1)$  or, equivalently, when the codimension r = N - n it gives r = n(n - 1)/2.
Furthermore, one motivation in the treatment of embedding has been devoted, namely,
Randall- Sundrum braneworld  model  is based on the assumption that four dimensional
space-time is a three-brane, embedded in a five-dimensional Einstein space \cite{Randall}.
On the other hand Nash in 1956 \cite{nash} established the idea of global isometric
embedding theorem of $V_n$  into Euclidean space $E_N$.

In a sense one could say that all Riemannian manifold has a local and a
global isometric embedding in an Euclidean space.  This result opened new perspectives
in embedding theorems with increasing degrees of generality and soon it became an
powerful tool to construct and classify solutions of GR. Inspired by these advances, a popular approach has been emerged in embedding 4-dimensional space-time into 5-dimensional flat space-time by using the spherical coordinates transformation and known as embedding class one if it satisfies the Karmarkar condition \cite{Karmarkar}.
One of the primary motivations of such embedding is to establish a relationship
between metric potentials and obtained exact solutions of Einstein's equation
to a single-generating function. In connection to this an exact anisotropic solution
of embedding class one, has been developed in \cite{Maurya(2016),Maurya2017a,Newton2016}. They showed that
in seeking solution for relativistic static fluid spheres one can utilized this technique
successfully. Motivated by the above facts, we will see how this theorem can be utilized for
obtaining an exact solution of Einstein's field equation for compact star models.
In this article we consider the static spherically symmetric spacetime metric with
embedding class one conditions, which can be altered to fit with a set of astrophysical objects.

 This paper is outlined in the following manner. In Sec. II, we present the structural equations
for anisotropic fluid distributions of stellar models applying the embedding class one conditions.
Specific models are then analyzed in a brief description by obtaining a
particular form of mass function. In Section III, we match our interior solution to an exterior
Schwarzschild vacuum solution at the boundary surface and then determine the
values of constant parameters. In Section IV, we continue our
discussion thorough geometrical analysis of the solution such as energy conditions,
hydrostatic equilibrium and stability of the star by fixing of certain parameters. We end
our discussion by concluding remarks in V.

\section*{II.~~Class one condition for Spherical symmetric metric and General relativistic equations}

The simplest configuration for a star is the static and spherically symmetric geometry has the usual form

\begin{equation}
ds^{2}=e^{\nu(r)}dt^{2}-e^{\lambda(r)}dr^{2}-r^{2}\left(d\theta^{2}+\sin^{2}\theta d\phi^{2} \right),\label{eq1}
\end{equation}
where the coordinates (t, r, $\theta$, $\phi$) are the spherical coordinates and the metric
coefficient $\lambda$ and $\nu$ are the functions of the radial coordinate $r$, and yet
to be determined by solving the Einstein equations.

As the metric (1) is time independent and spherically symmetric, we restrict ourselves that
the space time is of emending class one, if it
satisfies the Karmarker condition (see Ref. \cite{Karmarkar,Maurya(2016)} for more details discussion) and
metric functions $e^{\lambda}$ and $e^{\nu}$ are dependent on each other as:
\begin{equation}
e^{\lambda}=1+K\,F'^2,  \label{eq2}
\end{equation}
where \ $F=e^{\nu/2}$, $F = F(r)$ and $K > 0$. However $F$ = constant, leads to flat space time metric.

The energy momentum tensor $T_{\mu\nu}$ associated to a spherical distribution of matter bounded
by gravitation is locally anisotropic, that is $T_{\mu\nu}$ = $\text{diag} \left(\rho, -p_r, -p_t, -p_t\right)$,
where $p_r$ and $p_t$
are the radial and tangential pressures and $\rho$ is the energy density of the fluid, respectively.
Thus, the Einstein field equation, $G_{\mu\nu}$ = 8$\pi T_{\mu\nu}$,
where $G_{\mu\nu}$ is the Einstein tensor then reduce to the following ordinary differential equations
for the metric \label{eq1} (since we use natural units where G = c = 1) as:
\begin{equation}
 8\pi\,\rho(r) = e^{-\lambda}\,\left[\frac{\lambda'}{r}+\frac{e^{\lambda}-1}{r^2}\right],  \label{eq3}
\end{equation}
where the primes ($\prime$) denote differentiation with respect to r.
Then one may write the solution in a very familiar form of mass function of the compact object,
\begin{equation}
m(r)=\int_{0}^{r}{4\,\pi\,r^2\,\rho(r)\,dr}. \label{eq4}
\end{equation}
It is worth noting that the mass is the density inside a proper volume element
within a radius r. Using the above Eq. (\ref{eq4}) and Eq. (\ref{eq2}), we can write
the metric component $e^\lambda$  which is given by the equality
\begin{equation}
e^{-\lambda}=1-\frac{2m}{r}=(1+K\,F'^2)^{-1}. \label{eq5}
\end{equation}
Rewriting the above expression in terms of mass function as:
\begin{equation}
m(r)= \frac{K\,r\,F'^2}{2\,(1+K\,F'^2)} \,\,\, ~~\text{or}~~\,\, F'^2=\frac{2m}{K\,(r-2m)}. \label{eq6}
\end{equation}
	
Since $F'(0) = 0$ and $F^{\prime\prime}(0)> 0$ because $F=e^{\nu/2}$, $\nu^{\prime}(0)= 0$
and $\nu^{\prime\prime}(0)> 0$  which is already proved by Herrera \emph{et al.} \cite{Herrera1}
and Maurya \emph{et al.} \cite{Maurya1}.
From Eq. (\ref{eq6}), we obtain $m(0)\,= \,m'(0)\,=\, m''(0)=0$ and $m'''(0)=3\,F''(0)/K > 0$.
This solution clearly implies that $m(r)$ is  positive and  monotonic increasing function of r.

It is possible to derive the mass function in terms of F(r). Aiming this
the value of F(r) from Eq. (\ref{eq6}), we have
 \begin{equation}
 F(r)=\frac{1}{\sqrt{K}}\,\int{\sqrt{\frac{2m}{r-2m}}}. \label{eq7}
 \end{equation}
To solve this integral, we shall assume $2m(r) = r-f(r)$, that we claim $f(r)$
should be zero at centre i.e. $f(0)=0$ due to m$(0)=0$, which indicating that
the mass function is regular at the origin. Equation (\ref{eq7}) can be rewritten as
 \begin{equation}
 F (r) = \frac{1}{\sqrt{K}}\,\int{\sqrt{\frac{r-f(r)}{f(r)}}}. \label{eq8}
 \end{equation}

To make the above integral equations tractable and construct a physically viable model,
we suppose that $\frac{r-f(r)}{f(r)}=ar^4+br^2$  in a particular form, which gives
\begin{equation}
 f(r)=\frac{r}{ar^4+br^2+1}. \label{eq9}
\end{equation}
where two constants a, b are positive and makes the Eq. (\ref{eq8}) in a
simplest form
 \begin{equation}
F (r) = \frac{1}{\sqrt{K}}\,\int{\sqrt{a\,r^4+b\,r^2}}. \label{eq10}
 \end{equation}
Solving the integral gives
\begin{equation}
 F=A+\frac{(a\,r^2+b)^{3/2}}{3\,a\,\sqrt{K}}, \label{eq11}
 \end{equation}
where A stands for integrating constant. Substituting Eq. (\ref{eq11}) in Eq. (\ref{eq2}),
we obtain one of the metric coefficient in the form
\begin{equation}
 e^{\lambda}=1+b\,r^2+a\,r^4, \label{eq12}
 \end{equation}
and the mass function is given by
\begin{equation}
 m(r)=\frac{r^3\,(a\,r^2+b)}{2\,\Bigl(1+b\,r^2+a\,r^4 \Bigl)}. \label{eq13}
\end{equation}
 The motivation of choice for particular dimensionless function lies on the fact that
the obtained mass function a monotonic decreasing energy density in the interior of the star.
To construct a physically viable model this type of mass function is not new but similar works
 have been considered earlier by  Matese \& Whitman \cite{Matese} and Finch \& Skea \cite{Finch}
for isotropic fluid spheres, and Mak \& Harko \cite{Mak} for anisotropic fluid spheres.

 The system of equations used to study for spherically symmetric configurations with anisotropic
fluid distribution are
\begin{eqnarray}
\hspace{-2.6cm}
8\pi\,\rho (r)=\frac{b^2\,r^2 + a\,r^2\, (5 + a\,r^4) + b \Bigl(3 + 2\,a\,r^4\Bigl)}{\Bigl(1 + b\,r^2 + a\,r^4\Bigl)^2}, \label{eq14}
\end{eqnarray}
\begin{equation}
8\pi\,p_r (r)= -\frac{1}{\Bigl(a\,r^4+b\,r^2+1\Bigl)}\left[{b+a\,\left(r^2-\frac{6\,B\,\sqrt{a\,r^2+b}}
{A+B\,(a\,r^2+b)^{3/2}}\right)}\right],  \label{eq15}
\end{equation}
\begin{equation}
8\pi\,p_t(r) = \frac{a^2\,r^2\,B\,(a\,r^4+b\,r^2+9)-b\,(b^2\,B+A\,\sqrt{a\,r^2+b})+f_1(r)}{\sqrt{a r^2+b}~~ \Bigl(1 + b r^2 + a r^4\Bigl)^2~~ \Bigl[A + B (b + a r^2)^{3/2}\Bigl]}, \label{eq16}
\end{equation}
where we use $f_1(r)=a\,\left[6\,b\,B-b^2\,B\,r^2-2\,A\,r^2\,\sqrt{ar^2+b}\right]$ for our notational
convention.

Firstly, we will present the anisotropic effect by a term $(p_t - p_r)/r$,
 by taking into account Eqs. (\ref{eq15}-\ref{eq16}),
 may be expressed in the following equivalent form
\begin{eqnarray}
 \Delta=\frac{r^2\,\Bigl[B\,(b^2+a^2\,r^4)+A\,g - a\,B\,(3-2\,b\,r^2)\,(b^2+a^2\,r^4-a+2\,a\,b\,r^2)\,\Bigl]}{8\pi\,\sqrt{a\,r^2+b}~~ \Bigl(1 + b \,r^2 + a\,r^4\Bigl)^2~~
 \Bigl[A + B (b + a r^2)^{3/2}\Bigl]}, \label{eq17}
\end{eqnarray}
which representing a force that is due to the anisotropic nature of the fluid. The anisotropy will be
repulsive or directed outwards if $p_t > p_r$, and attractive or directed inward when $p_t < p_r$.
The effects of anisotropy forces maintain the stability and equilibrium configurations
of a stellar stricture, as we discuss later.

\begin{figure}[h!]
\centering
\includegraphics[width=5cm]{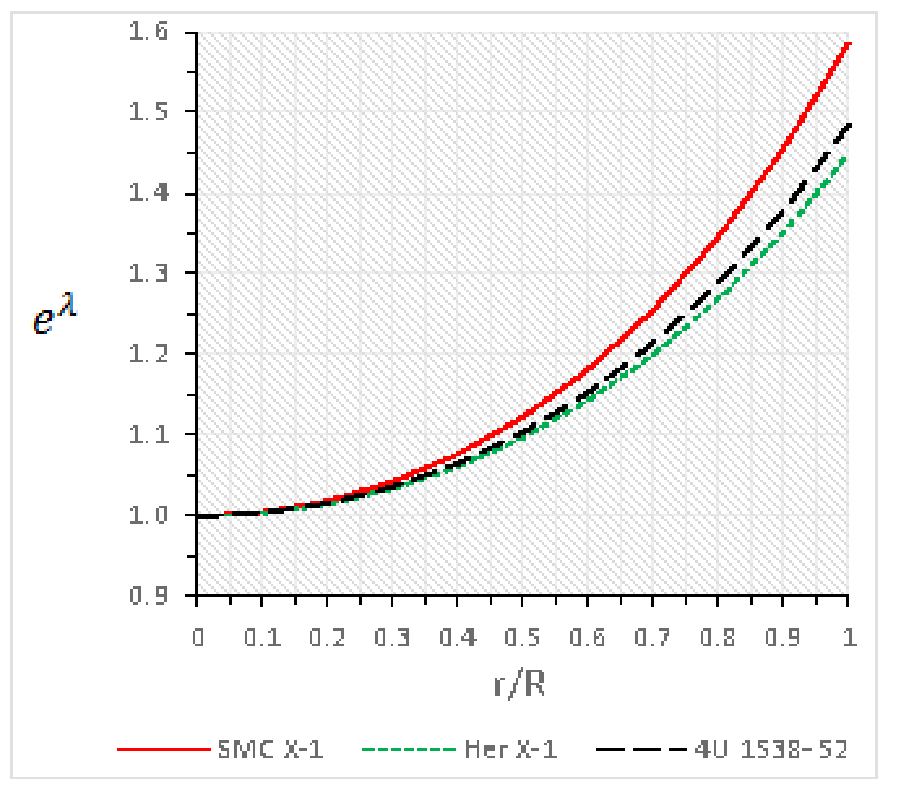} \includegraphics[width=5cm]{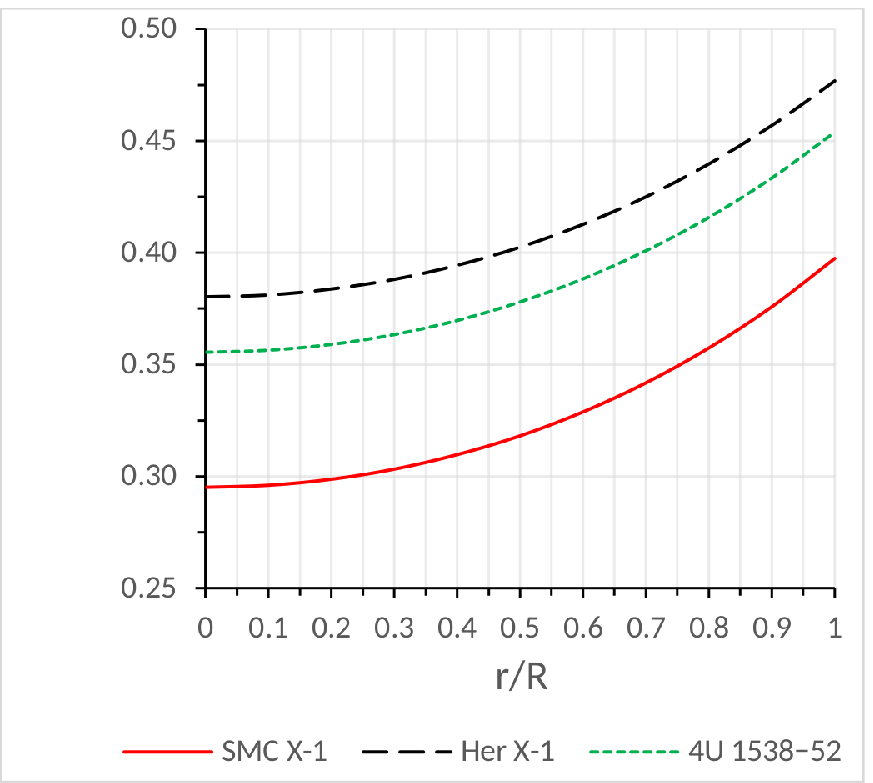}
\caption{\emph{Variation of the metric coefficient $e^{\lambda}$ and $e^{\nu}$ are shown
 as a function of radial coordinate for compact stars candidates SMC X-1, Her X-1 and 4U 1538-52
 with their respective parameters given in Table 1. Note that, qualitatively, the nature
of $e^{\lambda}$ and $e^{\nu}$ are monotonic increasing towards the boundary}.}
\end{figure}

  In more detail, it is important to emphasize the results. That's why we extend our
calculation for the first order differential equation, which are
\begin{eqnarray}
\frac{d\rho}{dr}= - \frac{2\,r\,\left[\,a^3 r^8 + 3 a^2 r^4 (4 + b r^2) + b^2 (5 + b r^2) + a (-5 + 13 b r^2 + 3 b^2 r^4)\,\right]}{8\pi\,\Bigl(1 + b r^2 + a r^4\Bigl)^3},  \label{eq18}
\end{eqnarray}

\begin{eqnarray}
\hspace{-0.6cm} \frac{dp_r}{dr}= {\frac{-2ar\,(g\,f^2+9\,a\,B^2 g^{3}-3\,a\,B\,f)}{8\,\pi\,g~f^2\,\Bigl(1 + b\,r^2 + a\,r^4\Bigl)}} + \frac{2\,r\,(2\,a\,r^2+b)\,(f~g^2-6\,B\,g)}{8\,\pi\,\Bigl(1 + b\,r^2 + a\,r^4 \Bigl)^2~f}, \label{eq19}
\end{eqnarray}

\begin{eqnarray}
\hspace{-1.7cm} \frac{dp_t}{dr}= \frac{2\,r\,\Bigl[\,\Psi_1(r)+\Psi_2(r)+\Psi_3(r)+\Psi_4(r)+\Psi_5(r)\,\Bigl]}
{16\,\pi\,\,(b + a r^2)^{3/2} (1 + b r^2 + a r^4)^3 \Bigl[A + B (b + a r^2)^(3/2)\Bigl]^2}, \label{eq20}
\end{eqnarray}
for our notational conventional we use
$f=\Bigl[A + B (b + a r^2)^{3/2}\Bigl]$, ~~ $g=\sqrt{b+ a r^2}$, ~~~ and\\

  $\Psi_1(r)=-6 a^6 B^2 r^{12} g + 4 b^3 \Bigl[2 A b^2 B + A^2 g + b^3 B^2 g ] + a^5 B r^8 [15 A r^2 - 2 B g (44 + 9 b r^2)\Bigl]$,\\

  $\Psi_2(r)=2 a b \Bigl[-11 b^3 B^2 g h - 2 A^2 g (4h-3) - A b^2 B (21 h - 8)\Bigl]$,\\

  $\Psi_3(r)=-a^4 B r^4\,\Bigl[-7 A x (-8 + 9 b r^2) + 2 B g (9 + 131 b r^2 + 4 b^2 r^4)\Bigl]$,\\

  $\Psi_4(r)=a^2 \Bigl[4 A^2 r^2 g (-1 + 6 b r^2) - 2 b^2 B^2 g (3 + 65 b r^2 - 21 b^2 r^4) + 3 A b B (4 - 31 b r^2 f)\Bigl]$, \\

  $\Psi_5(r)=a^3 r^2 \Bigl[12 A^2 r^4 g - 2 b B^2 g [12 + b r^2 (141- 14 b r^2)] + A B (9 - 123 b r^2 + 107 b^2 r^4)\Bigl]$.

We exercised our results in various ways first by checking calculation and putting
restriction on the physical parameter  based on logarithmic principle
and then by using graphical representation which are illustrated in Figs. (1-3),
describing the metric functions, energy density, radial and transverse pressures
and measure of anisotropic within the given radius. In the next section we will
look for supplementary restrictions on the model to make it physically viable.

\begin{figure}[h!]
\centering
\includegraphics[width=5cm]{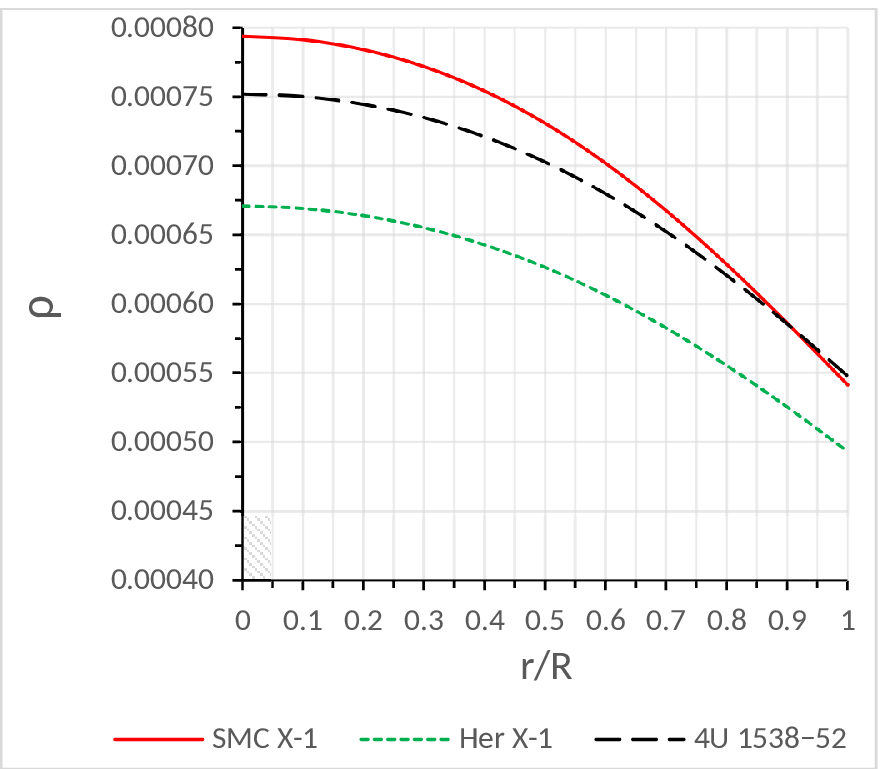} \includegraphics[width=5cm]{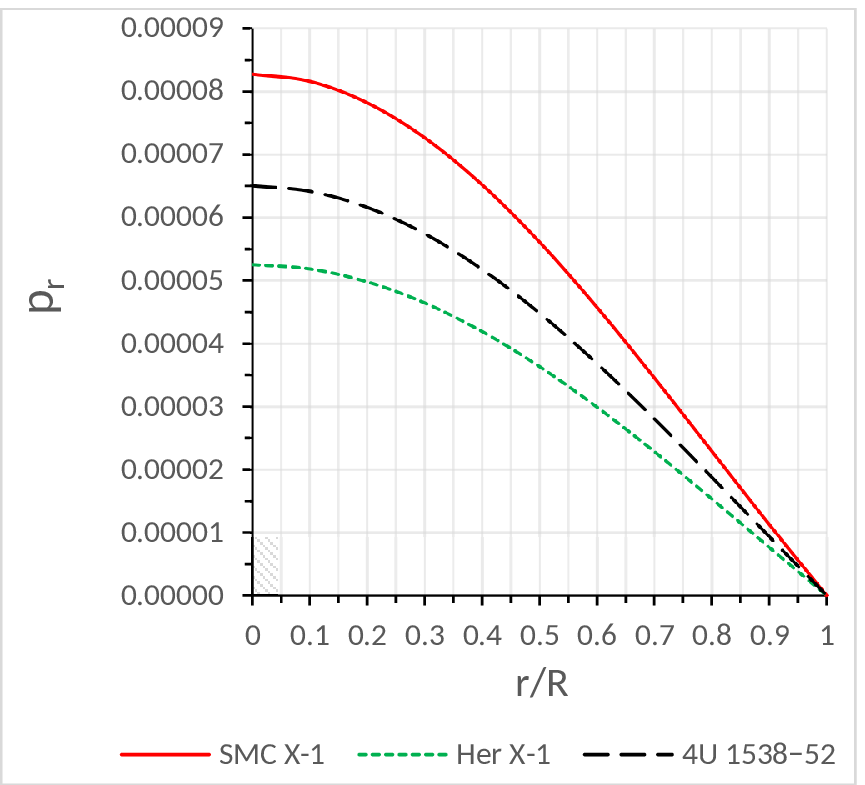} \\
\includegraphics[width=5cm]{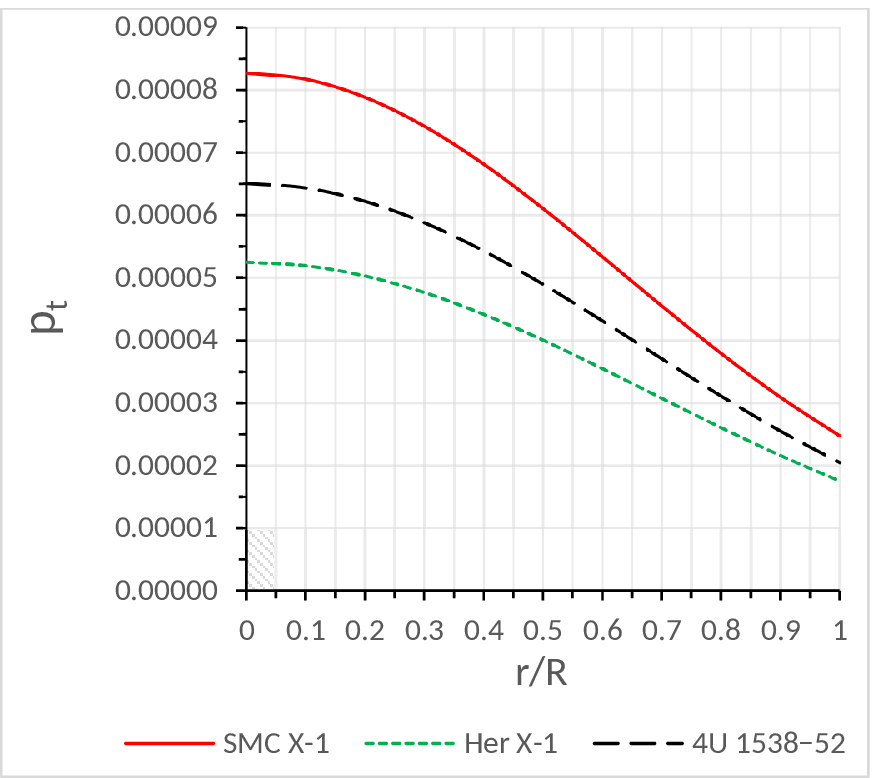}  \includegraphics[width=5cm]{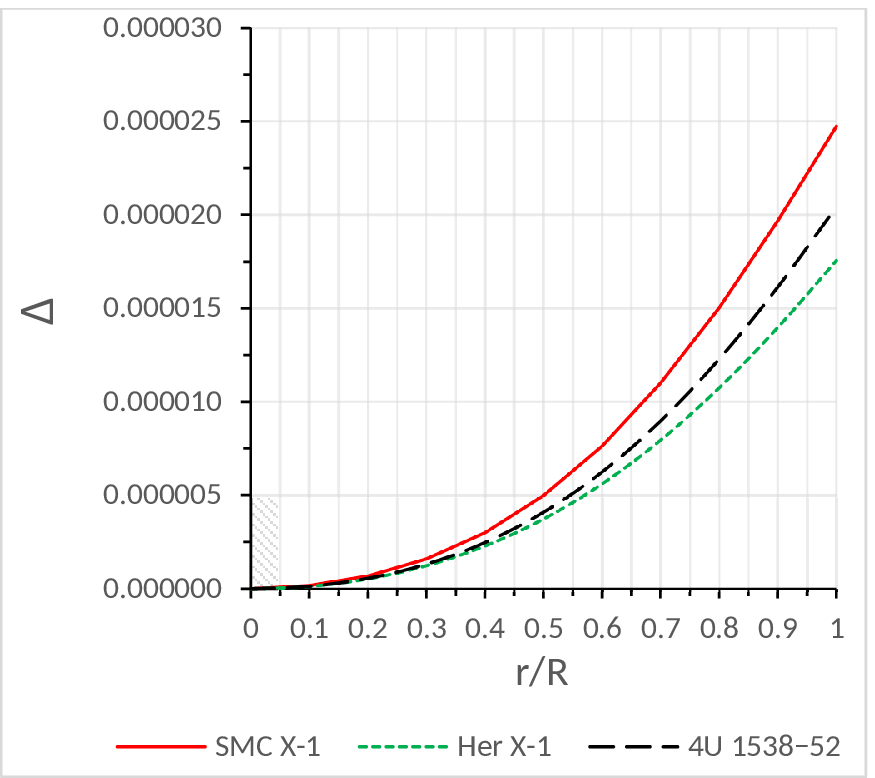}
\caption{ \emph{The energy density, radial pressure, transverse pressure and anisotropy factor in their normalized
forms as a function of the radial coordinate are shown according to Eq. (14-17). The pressure anisotropy is positive in the interior of star, i.e., force due to the anisotropic nature is directed outward.} }
\end{figure}

\section*{III.~~~Boundary conditions}

To do the matching properly, we start by joining an interior spacetime
$\mathcal{M_{-}}$, to an exterior $\mathcal{M_{+}}$, Schwarzschild vacuum solution
at the boundary surface r = R. At this boundary the metric should be continuous. The Schwarzschild
solution is given by
\begin{eqnarray}\label{21}
&\qquad\qquad\hspace{-2.7cm}ds^{2} =\left(1-\frac{2M}{r} \right)\, dt^{2} -\left(1-\frac{2M}{r}
\right)^{-1} dr^{2}-r^{2} (d\theta ^{2}
+\sin ^{2} \theta \, d\phi ^{2} ),\nonumber \\
\end{eqnarray}
where $M$ denote the total mass of the compact star. In deciding criterions
for an anisotropic compact star the radial pressure $p_{r}$ must be finite and positive
inside the stars, and it should vanishs at the boundary $r = R$ of the star~\cite{Misner}.

 In determining the value of constrains parameters we use the boundary condition
 $p_r(R) =0$, which can express as
\begin{eqnarray} \label{22}
\frac{A}{B}=\frac{\sqrt{b + a\, R^2}~ \Bigl[\,6 a B  - b^2 B - 2\,a\,b\, B\, R^2 -a^2\,B\, R^4 \,\Bigl]}{(b + a\,R^2)}. \label{16}
\end{eqnarray}
In order to match smoothly on the boundary surface, we must require the continuity of the first and the second fundamental forms across that surface. Then it follows the condition $e^{\nu(R)}=e^{-\lambda(R)}$
which gives the value of constant parameter B as
\begin{eqnarray}
B=\frac{1}{\sqrt{1+bR^2+aR^4}~\left[\,\frac{A}{B}+(aR^2+b)^{3/2}\,\right]},  \label{23}
\end{eqnarray}
and using the condition $e^{-\lambda(R)}=1-\frac{2M}{R}$, we have
\begin{eqnarray}
   M=\frac{R^3\,\left(aR^2+b \right)}{1+bR^2+aR^4}.  \label{24}
\end{eqnarray}
This represents the total mass of the sphere as seen by an outside observer.
Additionally, bounds on stellar structures is an important source of information
and classification criterion for compact objects, to determine the mass-radius ratio which
was proposed by  Buchdahl \cite{Buchdahl}. This bound has been considered for thermodynamically stable
perfect fluid compact star with ratio 2M/R, must be less than 8/9. We have carried
out the analysis for our compact star candidates SMC X-1, Her X-1 and 4U 1538-52,
that are used to calculate the values of constants A and B. For this purpose, we present these
results (central density, surface density, central pressure, mass-radius ratio) in table I \& II, respectively.
\begin{center}
\begin{tabular}{ |p{3cm}||p{2cm}|p{2cm}|p{2cm}|p{2cm}|p{2cm}|p{2cm}| }
\hline Table~I & \multicolumn{6}{c|}{Numerical values of the parameters or constants for different compact stars ~\cite{Gangopadhyay}}\\
\hline Compact Star & ~~R~(km) & ~~M$\left(M_\odot\right)$ & $~~~a\,(km^{-4})$ & ~~~$b\,(km^{-2})$ & ~~~~A &~$B(km^{3})$ \\
\hline SMC X-1  & 8.301  & 1.04 & 0.0000270 & 0.006648 & 0.439167064 & 452.08447 \\
\hline Her X-1 & 8.1  & 0.85 & 0.0000185 & 0.005620 & 0.481332147 & 618.82624 \\
\hline 4U 1538-52 & 7.866 & 0.87 & 0.0000247 & 0.006300 & 0.481382726 & 490.02078 \\
\hline
\end{tabular}
\end{center}
  Let us now focus on surface gravitational redshift, which gives a wealth of information about
compact objects, and defined by Z = $\Delta \lambda/\lambda_{e} $ =
$\frac{\lambda_{0}-\lambda_{e}}{\lambda_{e}}$, where $\lambda_{e} $ is the emitted wavelength
at the surface of a nonrotating star and the observed wavelength $\lambda_{e}$. Thus
one can estimate the gravitational redshift, defined by $Z_s $, from the surface of
the star as measured by a distant observer by the following relation
\begin{equation}
Z_{s} = -1+\Bigl\rvert g_{tt}(R)\Bigl\rvert ^{-1/2} = -1+\left( 1-\frac{2M}{R}\right) ^{-1/2},
\end{equation}
where $g_{tt}$ = $e^{\nu(R)}$ = $\left( 1-\frac{2M}{R}\right)$.  Measurement of the
gravitational redshift for a static perfect fluid sphere is not larger than $Z_{s}$ = 2 \cite{Buchdahl},
whereas for an anisotropic fluid sphere this value may be increase up to $Z_{s}$ = 3.84, as given in Ref. \cite{Ivanov}.
We are trying to estimate the surface redshift given table II for different compact star candidates.

\begin{center}
\begin{tabular}{ |p{3cm}||p{2.5cm}|p{2.5cm}|p{2.5cm}|p{2.4cm}|p{2.2cm}|  }
\hline Table~II  & \multicolumn{5}{c|}{Estimated physical values based on the observational data and theory}\\ \hline
Compact star & $\rho_{0} (gm/cm^{3})$ & ~~$\rho_{R} (gm/cm^{3})$ & ~~$p_{c} (dyne/cm^{2})$ & ~~~ $2M/R$~~~& $~~~Z_s$\\ \hline
\hline SMC X-1 & 1.0710$\times 10^{15} $ & 7.3060$\times 10^{14} $ & 1.0052$\times 10^{35} $ & 0.36959 $<$ 8/9 & 0.25948 \\

\hline Her X-1 & 9.0538$\times 10^{14} $ & 6.6539$\times 10^{14} $ & 6.3783$\times 10^{34}$ & 0.30957 $<$ 8/9 & 0.20348\\
\hline 4U 1538-52  & 1.0149$\times 10^{15} $ & 7.3928$\times 10^{14} $ & 7.9036$\times 10^{34}$ & 0.32628 $<$ 8/9 & 0.218346 \\
\hline
\end{tabular}
\end{center}

\section*{IV.~~~Physical features and Comparative study of the physical parameters
for compact star model}

To proceed  further discussion based on the obtained solution that must satisfy some
general physical requirements. In order to simplify the analysis and make the solution
more viable we explore some physical features of the compact star and carry out a
comparative study between the data of the model parameters  with a set of astrophysical
objects in connection to direct comparison of some strange/compact star candidates.

\subsection{Energy Conditions}
Let us first discuss a very simple but important features of a stellar model.
The study of energy conditions within the framework of GR is an essential part for studying the compact objects.
Here we examine the energy conditions, namely : (i) Null energy condition
(NEC), (ii) Weak energy condition (WEC) and (iii) Strong energy condition (SEC),
at all points in an interior of a star holds simultaneously, by the following inequalities
\begin{eqnarray}
\textbf{NEC:}~ \rho(r)-p_r \geq  0,\\
\textbf{WEC:}~ \rho(r)-p_r(r) \geq  0~~(WEC_r),~~~ \rho(r)-p_t(r) \geq  0~~(WEC_t),\\
\textbf{SEC:}~ \rho(r)-3p_r(r) \geq  0~~(SEC_1), ~~~\rho-p_r(r)-3p_t(r) \geq  0 ~~(SEC_2).
\end{eqnarray}
From the above inequalities we provide a graphical representation that one can easily justify the
nature of energy condition for three different compact objects in Fig. 3.
Due to the complexity of the expression, we only able to write down the above
inequalities and plotted the graphs against the above criterions.
In fact, all the energy conditions are not violated and well behaved in the stellar interior.

\begin{figure}[h]
\centering
\includegraphics[width=4.5cm]{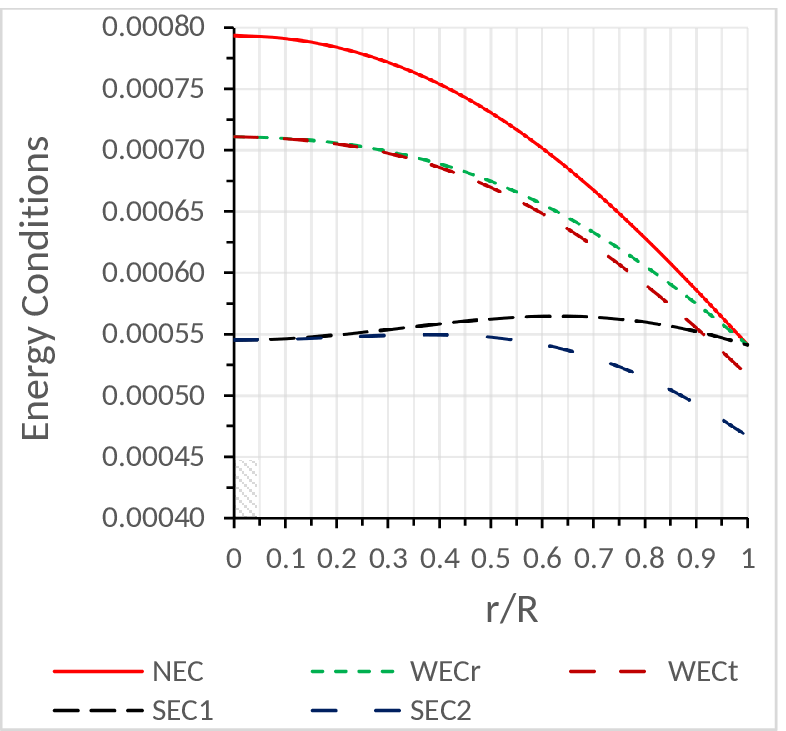} \includegraphics[width=4.5cm]{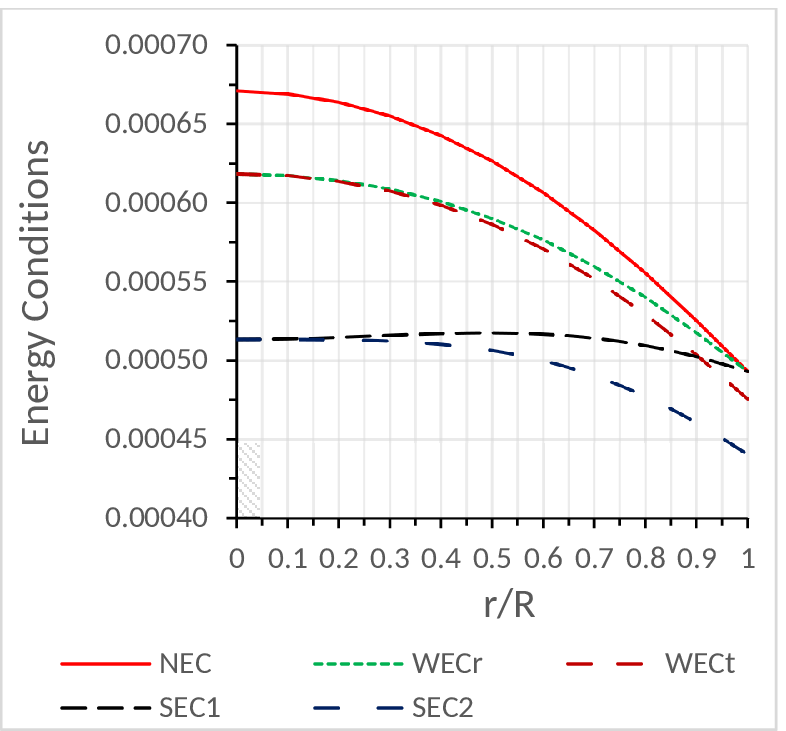} \includegraphics[width=4.5cm]{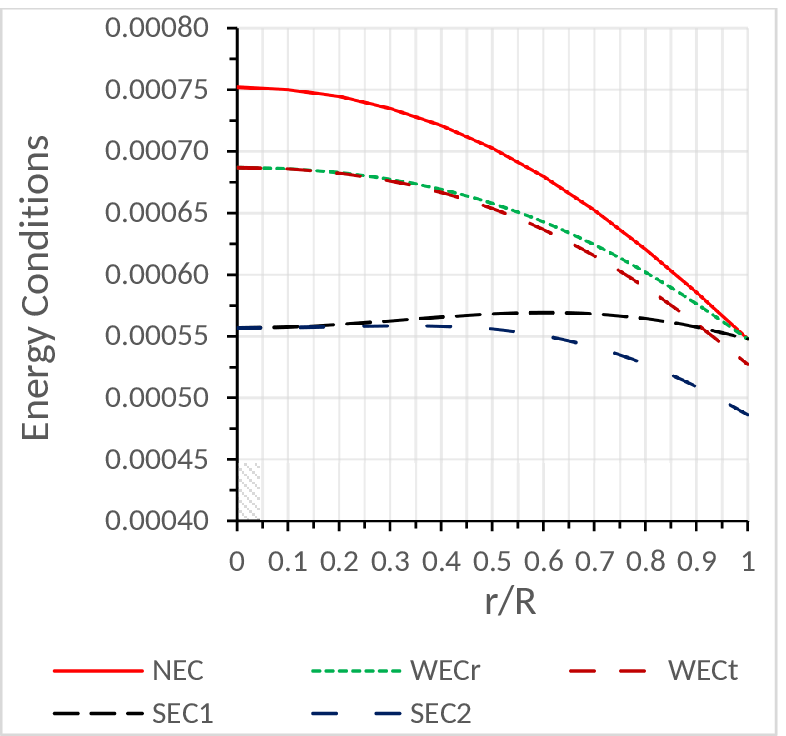}

\caption{ \emph{The standard energy conditions of GR have been plotted, namely, NEC, WEC and SEC
with the radial coordinate inside stellar structure for different compact star candidates
SMC X-1, Her X-1 and 4U 1538-52 from left to right. In every cases it satisfies all energy
conditions and the parameter values have been taken for the graphs from Table 1.} }
\end{figure}

\subsection{Generalized Tolman-Oppenheimer-Volkov Equation}
On the other hand, for a given compact star, it is possible to test for hydrostatic equilibrium
under the different forces namely gravitational, hydrostatic and anisotropic forces.
Nevertheless, in our investigations we need to apply
generalized Tolman-Oppenheimer-Volkov (TOV) equation \cite{Oppenheimer1939,Leon1993},
which is commonly used  for an anisotropic fluid distribution is given by
\begin{eqnarray}
& \qquad\hspace{-1cm}-\frac{M_G(r)(\rho+p_r)}{r^2}e^{\frac{\lambda-\nu}{2}}-\frac{dp_r}{dr}+\frac{2}{r}(p_t-p_r)=0, \label{eq25}
\end{eqnarray}
where the effective gravitational mass $M_G(r)$ is defined by
\begin{eqnarray}
&\qquad\hspace{1cm} M_G(r)=\frac{1}{2}{{r}^{2}}e^{\frac{\nu-\lambda}{2}}\nu'. \label{eq26}
\end{eqnarray}

Then equation (\ref{eq25}) may be rewritten as
\begin{eqnarray}
&\qquad\hspace{-1cm}-\frac{\nu'}{2}(\rho+p_r)-\frac{dp_r}{dr}+\frac{2}{r}(p_t-p_r)=0, \label{eq27}
\end{eqnarray}
Let us now attempt to explain the Eq. (\ref{eq27}) from an equilibrium point of view, which
was first shown by Tolman~\cite{Tolman1939} and Oppenheimer and then Volkoff~\cite{Oppenheimer1939},
where they predicted the stable equilibrium condition for the compact star as a sum of
three different forces, viz. gravitational force ($F_g$), hydrostatics force ($F_h$) and
anisotropic force ($F_a$). Thus the above condition assumed the following form, namely
\begin{equation}
F_g+F_h+F_a=0. \label{eq28}
\end{equation}
The components of forces can be expressed in explicit form as:
\begin{eqnarray}
&\quad\hspace{-1.9cm} F_g=  \frac{-\nu'(\rho+p_r)}{2}
=-\frac{6ar\,B\,g}{8\,\pi}\,\left[\frac{A\,(b + 2 a r^2) + B\, g~ [b^2 + 5 a^2 r^4 + a\, (3 + 6 b r^2)]} {(1 + b r^2 + a r^4)^2\, [A + B\, (b + a r^2)^{3/2}]^2}\right],\label{eq29}
\end{eqnarray}

\begin{eqnarray}
&\quad\hspace{-0.4cm} F_h=-\frac{dp_r}{dr}
=\frac{a\,r}{(1 + b\,r^2 + a\,r^4)}{\,\left[ \frac{gf^2+9\,a B^2 g^{3}-3\,a\,B\,f}{4\,\pi\,g~~f^2}\right]}  -\frac{r\,(2\,a\,r^2+b)}{(1 + b\,r^2 + a\,r^4)^2} \left[\frac{g^2\,f-6\,B\,g}{4\,\pi\,f}\right], \label{eq30}
\end{eqnarray}

\begin{eqnarray}
 &\quad\hspace{-0.5cm} F_a=\frac{2}{r}(p_t-p_r)
=\frac{r\,[B\,(b^2+a^2\,r^4)+A\,\sqrt{a\,r^2+b} - a\,B\,(3-2\,b\,r^2)\,(b^2+a^2\,r^4-a+2a\,b\,r^2)\,]}{4\,\pi\,\sqrt{a r^2+b}~~ \left(1 + b r^2 + a r^4\right)^2~~ \left[A + B (b + a r^2)^{3/2}\right]},  \label{eq31}
\end{eqnarray}
At this point in the derivation, a stable configuration of the system  is
counter balance by the components of different forces. To illustrate this in more
detail let us introduce an graphical representation as evidenced in Fig. 4.

Here, the combine forces of hydrostatic ($F_h$) and  anisotropic ($F_a$), dominate the
gravitational force ($F_g$),  while the anisotropic stress has a less role to the action of
equilibrium condition. Hence, following the reason outlined above and making
use of Eq. (\ref{eq28}), we established the stable configuration model as shown in Fig. 4.

\begin{figure}[h]
\centering
\includegraphics[width=4.5cm]{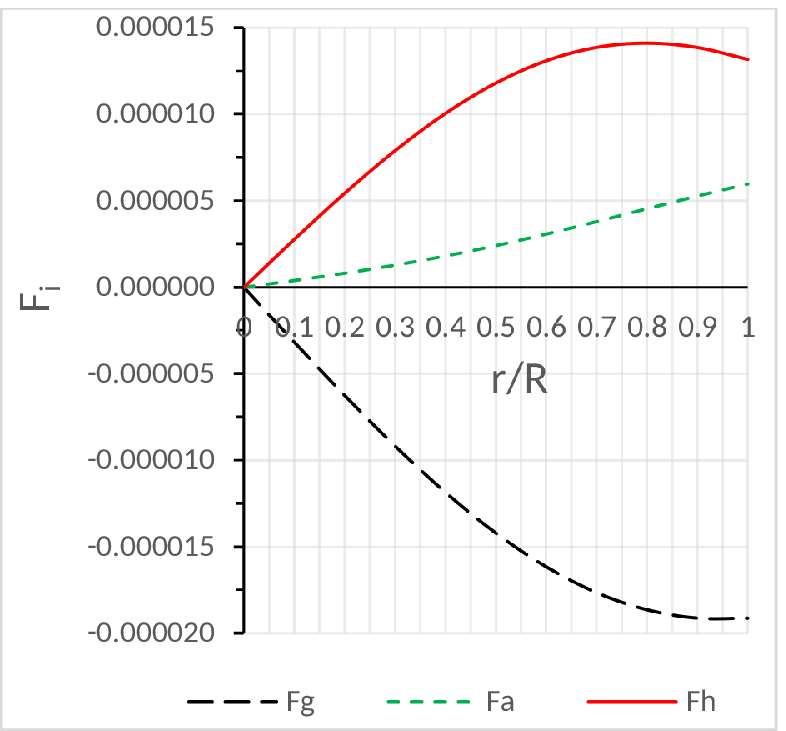} \includegraphics[width=4.5cm]{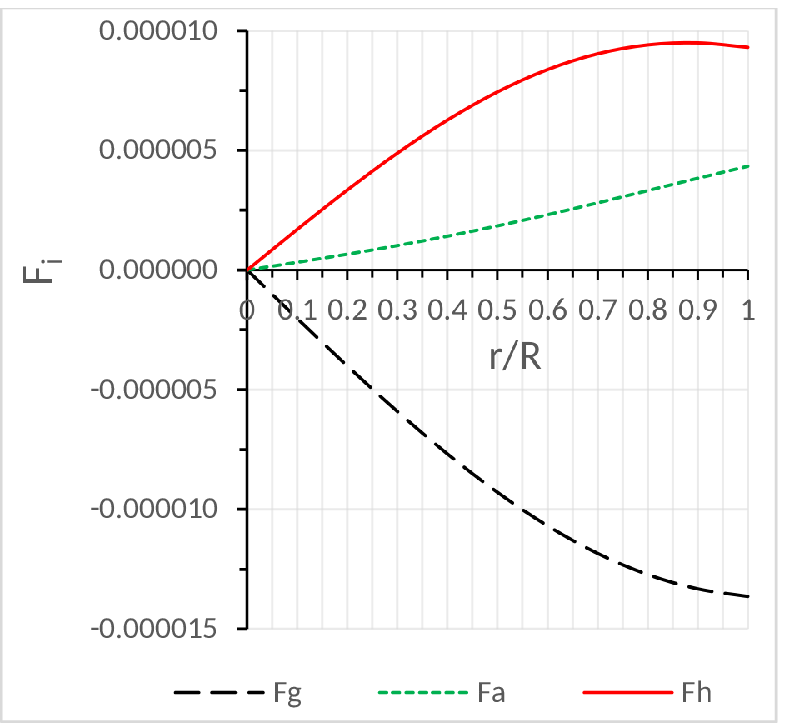}\includegraphics[width=4.5cm]{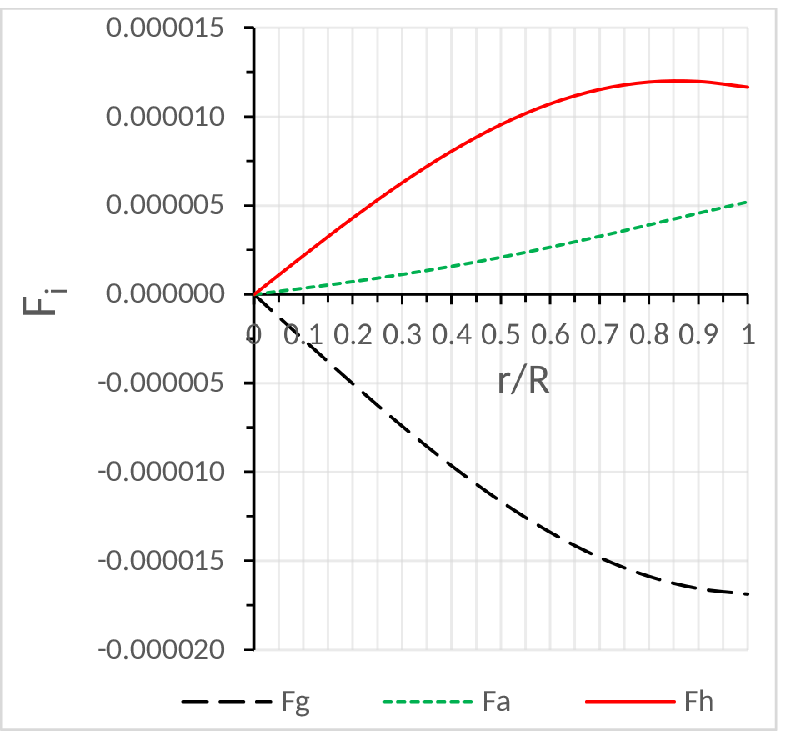}
\caption{\emph{Plots for the variation of different forces, namely, gravitational force ($F_g$),
hydrostatics force ($F_h$) and anisotropic force ($F_a$) with the radial coordinate
inside stellar structure. For compact star candidates SMC X-1, Her X-1 and 4U 1538-52,
different forces have been plotted from left to right. See the text for details about
how they maintain the stable configuration mode}.}
\end{figure}

\subsection{Stability Analysis}
In this section we analyze the speed of sound propagation $v^2_s $,  is given by the
relation $v^2_s = dp/ d\rho$. For a  physically interesting stellar geometry should
require that the sound speed does not exceed the speed of light, i.e., the physically
relevant region is always less than unity.  Here we will investigate the speed
of sound for anisotropic fluid distribution and propagating along radial as well as transverse direction,
should satisfy the bounds $0 < v_{r}^{2}=\frac{dp_r}{d\rho} < 1$ and $0 < v_{t}^{2}=\frac{dp_t}{d\rho} < 1$, \cite{Herrera(2016)}.

 We first carry out an analysis of sound velocity with graphical representation.
Our result for obtained velocity of sound for compact star with anisotropic matter is presented in Fig. 5,
for strange star candidates SMC X-1, Her X-1 and 4U 1538-52.
As the resulting expressions are very elaborated, so we only plot the results for different
compact stars. It is interesting to note that both $v_r^2,\,v_t^2 < 1$ and monotonic decreasing function,
which all show a behavior similar to that found for other compact objects. This is a
sufficient condition for the solution to be causal.
A notable characteristic of the sound velocity is the estimation of the potentially
stable and unstable eras, by considering the expression $0< \Bigl\rvert v_{t}^{2}-v_{r}^{2}\Bigl\rvert \leq 1$,
for stable potential \cite{Andreasson}. According to Fig. 5
(extreme right) our solution gives us the stable star configuration.

 Our investigation shows that obtained mass function by using the Karmarker condition
for strange compact star matter, satisfies both energy and stability conditions

\begin{figure}[h]
\centering
\includegraphics[width=4.6cm]{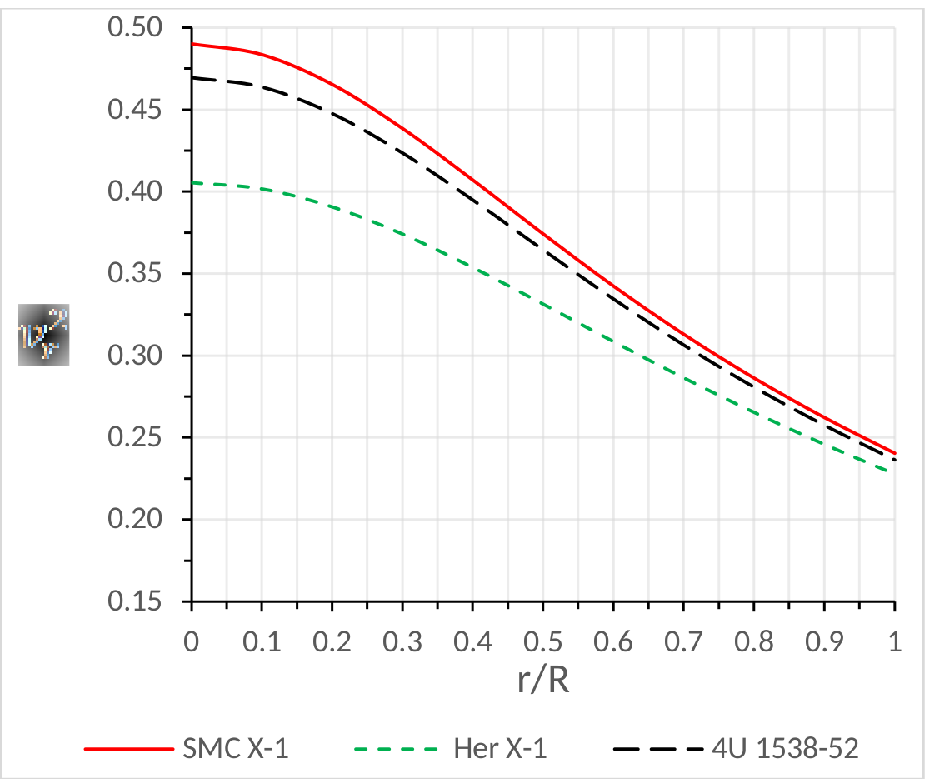} \includegraphics[width=4.6cm]{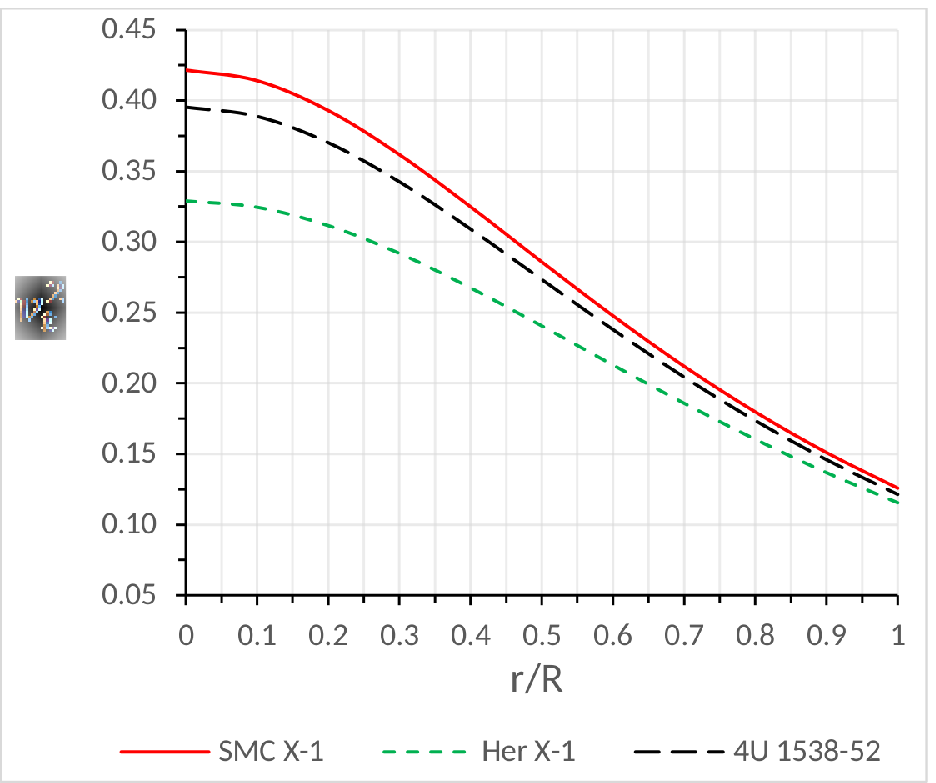} \includegraphics[width=4.5cm]{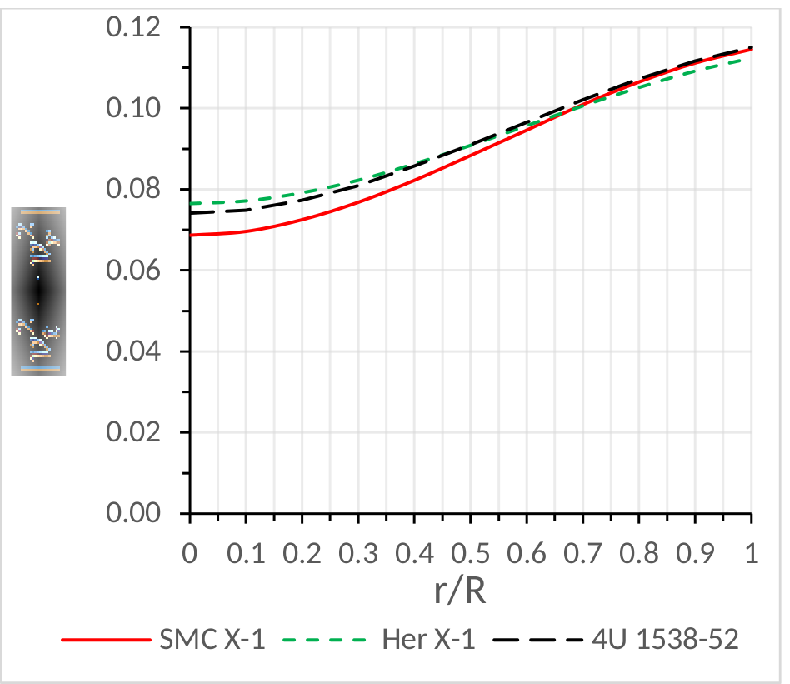}
\caption{\emph{Illustrative plots for speed of sound propagation $v_s^2 =d p/d\rho$ along the
radial and transverse direction for the same stars. For a view on the status it's clearly that
the speed of sound is less than unity. In extreme right position we have drawn the graphs for
$\Bigl\rvert v_{t}^{2}-v_{r}^{2}\Bigl\rvert $ using the same parameters enlisted in table 1.} }
\end{figure}

%%%%%%%%%%%%%%%%%%%%%%%%%%%%%%%%%%%%%%%%%%%%%%%%%%%%%%%%%%%%%%%%%%%%%%%%%%%%%%%%%

\section{Concluding Remarks}

To summarize, we have studied  compact objects supported by anisotropic fluid distribution,
which plays an important role in preventing gravitationally collapse. The motivation behind
such a construction is that,  we established a relation between metric potentials by imposing
an embedding theorem, known as embedding class one within the framework of GR. As a result of
this approach, we generate a mass function which we have used to study the interior of stellar
objects with density decreasing outwards. In next, we have started by deriving the basic
equations of Einstein’s field equations that describing the structure of compact objects.

In section (II), using the structural equations we derived the energy density,
radial and transverse pressures and measure the nature of anisotropy.  The complicated expressions
given by equations (14-17) are plotted as a function of the radius for our compact
stars candidates SMC X-1, Her X-1 and 4U 1538-52. An important feature of this model is the
energy density might not vanish at the boundary r =R,  though the radial pressure vanishes
for all parameter values in according to the boundary condition. Additionally, at the boundary
 the interior spacetime have been matched by a Schwarzschild metric, and determine the values
of arbitrary constants A and B for our compact stars candidates SMC X-1, Her X-1 and 4U 1538-52 in section (III).
A comparative study of our results with that of the compact star candidates
are provided in Table I and II.

Once the mass function is specified, in order to close the system based on physical requirements,
we further proceed by investigating the energy conditions, hydrostatic equilibrium
under the different forces, and velocity of sound. All the physical properties are well
behaved within the stellar radius. We also showed that upper bound of the mass-radius ratio must be less than 8/9 as proposed by Buchdahl \cite{Buchdahl}, for different compact star candidates which we
have used for our model.

This indicates that the approach adapted in this paper is likely to produce other meaningful models
with specific mass function that also greatly help in understanding the properties
of other different static compact configurations. In our further study it would be interesting to investigate
other forms of mass function that exhibit more general behaviour and thereby describe strange
stars related with observational details.

\section*{Acknowledgments} AB is thankful to the authority of
Inter-University Centre for Astronomy and Astrophysics, Pune,
India for providing research facilities.


\begin{thebibliography}{99}

\bibitem{Gangopadhyay} T. Gangopadhyay \emph{et al.}: \textcolor[rgb]{1.00,0.00,1.00}{{\it Mon. Not. R. Astron. Soc.}, {\bf 431}, 3216 (2013)}.
\bibitem{Lattimer} J. Lattimer: (2010) http://stellarcollapse.org/nsmasses.
\bibitem{Ivanov} B. V. Ivanov: \textcolor[rgb]{1.00,0.00,1.00}{{\it Phys. Rev. D}, {\bf 65}, 104011 (2002)}.
\bibitem{Witten} E. Witten: \textcolor[rgb]{1.00,0.00,1.00}{{\it Phys. Rev. D}, {\bf 30}, 272 (1984)}.
\bibitem{Glendenning} N. K. Glendenning, Ch. Kettner \& F. Weber: \textcolor[rgb]{1.00,0.00,1.00}{{\it Phys. Rev. Lett.}, {\bf 74}, 3519 (1995)}.
\bibitem{Shapiro} S. L. Shapiro \& S. A. Teukolosky: {\it Black Holes, White Dwarfs and Neutron Stars:
The Physics of Compact Objects} (Wiley, New York,1983).
\bibitem{Barreto} L. Herrera \& W. Barreto: \textcolor[rgb]{1.00,0.00,1.00}{{\it Phys. Rev. D.}, {\bf 88}, 084022 (2013)}.
\bibitem{Lai} X.Y. Lai \& R.X. Xu: \textcolor[rgb]{1.00,0.00,1.00}{{\it Astropart.Phys.}, {\bf 31}, 128-134 (2009)}.
\bibitem{Bowers}  R.L. Bowers \& E. P. T. Liang: \textcolor[rgb]{1.00,0.00,1.00}{{\it Class. Astrophys. J.}, {\bf 188}, 657 (1974)}.
\bibitem{Ruderman} R. Ruderman: \textcolor[rgb]{1.00,0.00,1.00}{{\it Rev. Astr. Astrophys.}, {\bf 10}, 427 (1972)}.
\bibitem{Gleiser} K. Dev \& M. Gleiser: \textcolor[rgb]{1.00,0.00,1.00}{{\it Gen. Rel. Grav.}, {\bf 34}, 1793, (2002)}.
\bibitem{Dev2003} K. Dev \& M. Gleiser: \textcolor[rgb]{1.00,0.00,1.00}{{\it Gen. Rel. Grav.}, {\bf 35}, 1435, (2003)}.
\bibitem{Herrera2002} L. Herrera, J. Martin \& J. Ospino: \textcolor[rgb]{1.00,0.00,1.00}{{\it J. Math. Phys.}, {\bf 43}, 4889, (2002)}.
\bibitem{Chaichian} M. Chaichian \emph{et al.}: \textcolor[rgb]{1.00,0.00,1.00}{{\it Phys. Rev. Lett}, {\bf 84}, 5261 (2000)}.
 \bibitem{Martinez} A. Perez Martinez, H. Perez Rojas \& H. J. Mosquera Cuesta: \textcolor[rgb]{1.00,0.00,1.00}{{\it Eur. Phys. J. C}, {\bf 29}, 111 (2003)}.
 \bibitem{Ferrer} E. J. Ferrer  \emph{et al.}: \textcolor[rgb]{1.00,0.00,1.00}{{\it Phys. Rev. C}, {\bf 82}, 065802 (2010)}.

\bibitem{Maurya2015a} S.K. Maurya  \emph{et al.}: \textcolor[rgb]{1.00,0.00,1.00}{{\it  Eur. Phys. J. C}, {\bf 75}, 389 (2015)}.
  \bibitem{Maurya2016a} S.K. Maurya  \emph{et al.}: \textcolor[rgb]{1.00,0.00,1.00}{{\it Astrophys. Space Sci.}, {\bf 361}, 163 (2016)}.

\bibitem{Hossein} Sk. Monowar Hossein  \emph{et al.}: \textcolor[rgb]{1.00,0.00,1.00}{{\it  Int.J.Mod.Phys. D}, {\bf 21}, 1250088 (2012)}.
\bibitem{Mehedi2012} Mehedi Kalam  \emph{et al.}: \textcolor[rgb]{1.00,0.00,1.00}{{\it Eur. Phys. J. C}, {\bf 72}, 2248 (2012)}.

\bibitem{Mak(2002)} M.K. Mak \& T. Harko: \textcolor[rgb]{1.00,0.00,1.00}{{\it Chin. J. Astron. Astrophys.,}, {\bf 2}, 248, (2002)}.
\bibitem{Sharma} R. Sharma \& S. D. Maharaj: \textcolor[rgb]{1.00,0.00,1.00}{{\it Mon. Not. R. Astron. Soc.}, {\bf 375}, 1265 (2007)}.
\bibitem{Varela} Victor Varela  \emph{et al.}: \textcolor[rgb]{1.00,0.00,1.00}{{\it  Phys.Rev. D}, {\bf 82}, 044052 (2010)}.
\bibitem{Riemann} B. Riemann \& Abh. K$\ddot{o}$nigl: \textcolor[rgb]{1.00,0.00,1.00}{{\it gesellsch.}, {\bf 13}, 1 (1868)}.

\bibitem{Schlafli} L. Schl$\ddot{a}$: \textcolor[rgb]{1.00,0.00,1.00}{Ann. di Mat. $2^e$ s$\acute{e}$rie {\bf 5}, 170 (1871)}.
\bibitem{Randall} L. Randall \& R. Sundrum: \textcolor[rgb]{1.00,0.00,1.00}{{\it Phys. Rev. Lett.}, {\bf 83}, 3370 (1999)};
:\textcolor[rgb]{1.00,0.00,1.00}{ {\it Phys. Rev. Lett.}, {\bf 83}, 4690 (1999)}.

\bibitem{nash} J. Nash: \textcolor[rgb]{1.00,0.00,1.00}{{\it Ann. Math.}, {\bf 63}, 20 (1956)}.
\bibitem{Karmarkar} K. R. Karmarkar: \textcolor[rgb]{1.00,0.00,1.00}{{\it Proc. Ind. Acad. Sci. A}, {\bf 27}, 56 (1948)}.
\bibitem{Maurya(2016)} S. K. Maurya  \emph{et al.}: \textcolor[rgb]{1.00,0.00,1.00}{{\it Eur. Phys. J. A}, {\bf 52}, 191 (2016)}.
\bibitem{Maurya2017a}S. K. Maurya \emph{et al.}: \textcolor[rgb]{1.00,0.00,1.00}{{\it Eur. Phys. J. C}, {\bf 77}, 45 (2017)}.
\bibitem{Newton2016} Ksh. Newton Singh \emph{et al.}: \textcolor[rgb]{1.00,0.00,1.00}{{\it Int.J.Mod.Phys. D}, {\bf 25}, 1650099 (2016)}.
\bibitem{Herrera(2016)} L. Herrera : \textcolor[rgb]{1.00,0.00,1.00}{{\it Phys. Lett. A}, {\bf 165}, 206 (1992)}.



\bibitem{Buchdahl} H. A. Buchdahl: \textcolor[rgb]{1.00,0.00,1.00}{{\it Phys. Rev.}, {\bf 116}, 1027 (1959)};\\
 H. A. Buchdahl: \textcolor[rgb]{1.00,0.00,1.00}{{\it Astrophys. J.}, {\bf 146} 275, (1966)}.
\bibitem{Ivanov}  B. V. Ivanov: \textcolor[rgb]{1.00,0.00,1.00}{{\it Phys. Rev. D}, {\bf 65}, 104011 (2002)}.

\bibitem{Matese} J. J.  Matese \& P. G. Whitman: \textcolor[rgb]{1.00,0.00,1.00}{{\it Phy. Rev. D}, {\bf 11}, 1270 (1980)}.
\bibitem{Finch} M. R. Finch  \& J. E. F. Skea: \textcolor[rgb]{1.00,0.00,1.00}{{\it Class. Quantum Grav.}, {\bf 6}, 467 (1989)}.
\bibitem{Mak} M. K.  Mak \& T. Harko: \textcolor[rgb]{1.00,0.00,1.00}{{\it Proc. Roy. Soc. Lond., A}, {\bf 459}, 393 (2003)}.


\bibitem{Herrera1} L. Herrera, J. Ospino \& A. Di Parisco: \textcolor[rgb]{1.00,0.00,1.00}{{\it Phys. Rev. D}, {\bf 77}, 027502 (2008)}.
\bibitem{Maurya1} S. K. Maurya, Y.K. Gupta \& S. Ray: arXiv: \textcolor[rgb]{1.00,0.00,1.00}{1502.01915 [gr-qc] (2015)}.
\bibitem{Misner} C. W. Misner \&  D. H. Sharp: \textcolor[rgb]{1.00,0.00,1.00}{{\it Phys. Rev. B}, {\bf 136}, 571 (1964)}.
\bibitem{Tolman1939} R.C. Tolman: \textcolor[rgb]{1.00,0.00,1.00}{{\it Phys. Rev.}, {\bf 55}, 364 (1939)}.


\bibitem{Oppenheimer1939} J.R. Oppenheimer \& G.M. Volkoff: \textcolor[rgb]{1.00,0.00,1.00}{{\it Phys. Rev.}, {\bf 55}, 374 (1939)}.
\bibitem{Leon1993} J. Ponce de Le{\'o}n: \textcolor[rgb]{1.00,0.00,1.00}{{\it Gen. Relativ. Gravit.}, {\bf 25}, 1123 (1993)}.

\bibitem{Varela2010} V. Varela: \textcolor[rgb]{1.00,0.00,1.00}{{\it Phys. Rev. D}, {\bf 82}, 044052 (2010)}.

\bibitem{Devitt1989} J. Devitt \& P.S. Florides: \textcolor[rgb]{1.00,0.00,1.00}{{\it Gen. Relativ. Gravit.}, {\bf 21}, 585 (1989)}.

\bibitem{Andreasson} H. Andreasson: \textcolor[rgb]{1.00,0.00,1.00}{{\it Commun. Math. Phys.}, {\bf 288}, 715 (2009)}.




\end{thebibliography}
 \end{document}